\documentclass[aps,prd,showpacs,preprintnumbers,nofootinbib]{revtex4}
\usepackage[dvips]{graphicx}
\usepackage{bm,latexsym,amsmath,amssymb,amsfonts}
\newcommand*{\D}{{\rm d}}
\newcommand*{\cH}{{\cal H}}
\begin{document}
\title{Gravitational waves from $p$-form inflation}

\author{Tsutomu~Kobayashi$^{1}$}
\email[Email: ]{tsutomu"at"gravity.phys.waseda.ac.jp}
\author{Shuichiro~Yokoyama$^{2}$}
\email[Email: ]{shu"at"a.phys.nagoya-u.ac.jp}
\address{\,\\ \,\\
$^{1}$ Department of Physics, Waseda University,
Okubo 3-4-1, Shinjuku, Tokyo 169-8555, Japan\\
$^{2}$ Department of Physics and Astrophysics, Nagoya University, Aichi 464-8602, Japan}

\begin{abstract}
Recently it was shown that an inflationary background can be realized
by any $p$-form field non-minimally coupled to gravity.
In this paper, we study gravitational waves generated during $p$-form inflation.
Even though the background evolution is identical to that in
conventional scalar field inflation, the behavior of gravitational waves
is different in $p$-form inflation.
In particular, we find that the propagation speed of gravitational waves differs from unity
in 2- and 3-form inflationary models.
We point out that
the squared speed becomes negative in the large field models.
The small field models are free from pathologies and the correction to
the spectrum of gravitational waves turns out to be very small.
\end{abstract}

\preprint{WU-AP/297/09}
\maketitle

\section{Introduction}

Our universe is well described by a homogeneous and isotropic
Friedmann-Lema\^{\i}tre-Robertson-Walker
background with small fluctuations on it.
This picture is supported by the observations of
smoothly distributed large scale structure in the universe
such as the cosmic microwave background (CMB) radiation. 
A quasi-de Sitter expansion at early times, i.e.,
inflation, is a very attractive paradigm to account for
the universe which is homogeneous and isotropic to a high degree,
and hence the inflationary paradigm is widely accepted.
One or more scalar fields (inflatons) are commonly believed to be
responsible for inflation, but it is still unclear what inflatons really are.
Therefore, it is important to discuss alternative inflationary models
that rely not on scalar fields but on other fields.
One can even radically argue that
since no scalar fields have been discovered in nature,
it might be natural to consider inflation driven by other fields!
Indeed, an inflationary model driven by vector fields was proposed recently~\cite{V1}
(see also earlier works~\cite{Ford, AP} and other related
papers~\cite{Bambam, Koivisto, Dimopoulos, Yoko1, Yoko2, Watanabe, Koh2, Koh}).
The essential ingredients of vector inflation are
a large number of randomly oriented vector fields and their non-minimal coupling to gravity.
A large number of fields are used to make the model compatible with
isotropy of the background.
The non-minimal coupling to gravity is required in order to realize slow-roll inflation.
Later, it was shown that inflation can be driven by any $p$-form field,
with scalar and vector inflation being the special cases with $p=0$ and $p=1$, respectively~\cite{p-form}.

In this paper, we study the behavior of gravitational waves generated during $p$-form inflation.
Gravitational waves are in general produced quantum mechanically in the
quasi-de Sitter stage of the early universe~\cite{St}.
They can be observed indirectly via imprints on the CMB
and directly by future detectors such as LISA~\cite{lisa} and DECIGO/BBO~\cite{decigo}.
Therefore, the inflationary gravitational wave is a powerful probe into the early universe.
Gravitational waves from 1-form inflation were already investigated in~\cite{V2}.

This paper is organized as follows.
In the next section we give a brief review of $p$-form inflation.
Then, in Sec.~III we consider tensor perturbations (gravitational waves)
and derive the actions governing their behavior on the $p$-form inflationary background.
We quantize the derived actions in Sec.~IV.
Finally we draw our conclusions in Sec.~V.
Calculation details are presented in Appendix.

\section{$p$-form inflation}

We begin with a brief review of a $p$-form inflationary background.
The metric is given by
\begin{eqnarray}
\D s^2=a^2(\eta)\left(
-\D\eta^2+\delta_{ij}\D x^i\D x^j
\right).
\end{eqnarray}
The first example is vector (i.e., 1-form) inflation proposed in~\cite{V1}.
Apparently, vector fields are incompatible with isotropy
of background cosmology because they induce off-diagonal spatial components
of the energy-momentum tensor.
However, one can evade this problem by invoking three mutually orthogonal vector fields
or a large number $N$ of randomly oriented fields.
The former case was first analyzed in the context of dark energy~\cite{AP}.
In the latter case, anisotropy is statistically suppressed~\cite{V1}.
The action for vector inflation driven by a large number of fields is given by
\begin{eqnarray}
S=\int \D^4 x\sqrt{-g}\left\{
\frac{R}{2\kappa^2}+\sum_{a=1}^{N}\left[-\frac{1}{4}F_{\mu\nu}^{(a)2}-V(I^{(a)})+\frac{1}{12}RI^{(a)}\right]
\right\},
\label{action_vector}
\end{eqnarray}
where $I^{(a)}:=A_\mu^{(a)2}$ and $F_{\mu\nu}^{( a)}:=\nabla_\mu A_{\nu}^{( a)}-\nabla_\nu A_\mu^{( a)}$.
Note here that the vector fields are non-minimally coupled to gravity.
Without this coupling the vector fields would have an effective mass term of order the Hubble rate, $H$,
which makes it difficult to realize slow-roll inflation.
The non-minimal coupling in Eq.~(\ref{action_vector}) cancels this contribution, and
the equation of motion for each $B_i^{(a)}:=A_i^{(a)}/a$ reduces to
the equation for a (minimally coupled) scalar field~\cite{V1}.
(The equation of motion also implies $A_0^{(a)}=0$.)
Let us assume that $N$ fields all have the magnitude of order $B$ initially (i.e., $B_i^2=B^2$).
Then, the energy-momentum tensor is given by
$T_0^{\;0}\simeq-N[(B')^2/2a^2+V]$ and $T_i^{~j}\simeq N[(B')^2/2a^2-V]\delta_i^{~j}$,
where a prime stands for the derivative with respect to $\eta$. 
In deriving the expression for the energy-momentum tensor we used the formula
\begin{eqnarray}
\sum_{a=1}^NB_i^{(a)}B_j^{(a)}\simeq \frac{N}{3}B^2\delta_{ij}+{\cal O}(1)\sqrt{N}B_iB_j,
\label{form_stat}
\end{eqnarray}
and omitted terms corresponding to the subleading contributions.
Thus, for a wide class of potentials we get an inflationary background which is
very similar to usual scalar field inflation, or, more precisely, what is called $N$-flation~\cite{Nflation}.

Recently, vector inflation was generalized to the cases with $p$-form fields in~\cite{p-form}.
2-form inflation is driven by a large number of 2-form fields and
can be described by the action
\begin{eqnarray}
S =\int\D^4 x\sqrt{-g}\left\{ \frac{R}{2\kappa^2}+\sum^{N}_{a=1}\left[-\frac{1}{12}F_{\mu\nu\rho}^{(a)2}-V(I^{(a)})+
\frac{R}{6} I^{(a)}+\frac{1}{2}A_{\mu\nu}^{(a)}R^{\nu}_{\;\rho}A^{\rho\mu}_{(a)}\right]\right\},
\label{2fac}
\end{eqnarray}
where $ I^{(a)}:=A_{\mu\nu}^{(a)2}$ and
\begin{eqnarray}
F_{\mu\nu\rho}^{(a)}:=\nabla_{\mu}A_{\nu\rho}^{(a)}+\nabla_{\rho}A_{\mu\nu}^{(a)}+\nabla_\nu A_{\rho\mu}^{(a)}.
\end{eqnarray}
The field equations for each $A_{\mu\nu}$ obtained from the action~(\ref{2fac})
imply $A_{0i}=0$ and
\begin{eqnarray}
B_i''+ 2\cH B_i'+4a^2 V_I B_i = 0,\label{bg:bi}
\end{eqnarray}
where $\cH:=a'/a$, and, instead of $A_{ij}$, we used the field $B_i$ defined by
\begin{eqnarray}
A_{ij} = a^2\varepsilon_{ijk}B_k(\eta)
\end{eqnarray}
with $\varepsilon_{ijk}$ being the totally antisymmetric symbol.
Noting that $I=2B_i^2$ for the background, one sees that Eq.~(\ref{bg:bi})
coincides with the equation of motion for a inflaton field.
We are considering a large number $N$ of 2-form fields so that
we may use the formula~(\ref{form_stat}) also in this case.
Assuming again that $N$ fields all have the magnitude of order $B$ initially (i.e., $B_i^2=B^2$),
we have the estimate
$T_0^{\;0}\simeq-N[(B')^2/2a^2+V]$, $T_0^{\;i}\simeq0$, and $T_i^{~j}\simeq N[(B')^2/2a^2-V]\delta_i^{~j}$,
where subleading terms are statistically suppressed.
We thus have the background Einstein equations
\begin{eqnarray}
3\cH^2 &=&\kappa^2N\left[ \frac{(B')^2}{2}+a^2V(I)\right],\\
2\cH'+\cH^2&=&-\kappa^2N\left[ \frac{(B')^2}{2}-a^2V(I)\right].
\end{eqnarray}
In the case of chaotic inflation, the potential is given by $V(I)=m^2 I/4$.

Let us move on to the case of the 3-form field. The action for 3-form inflation is
\begin{eqnarray}
S=\int\D^4 x\sqrt{-g}\left\{
\frac{R}{2\kappa^2}-\frac{1}{48}F_{\mu\nu\rho\sigma}^2-V(I)+\frac{1}{8}RI-
\frac{1}{2}A_{\mu\nu\rho}R^{\rho\sigma}A_\sigma^{~\mu\nu}
\right\},\label{3finf}
\end{eqnarray}
where $I=A_{\mu\nu\rho}^2$ and
\begin{eqnarray}
F_{\mu\nu\rho\sigma}:=\nabla_\mu A_{\nu\rho\sigma}-\nabla_\sigma A_{\mu\nu\rho}
+\nabla_{\rho}A_{\sigma\mu\nu}-\nabla_\nu A_{\rho\sigma\mu}.
\end{eqnarray}
Note that a 3-form field is compatible with spatial isotropy and so in this case
we do not need to use a large number of fields.
We write the ansatz as follows:
\begin{eqnarray}
A_{0ij}=\alpha_{ij}(\eta),\quad A_{ijk}=a^3\phi(\eta)\varepsilon_{ijk}.
\end{eqnarray}
Substituting this into the field equations for the 3-form field derived from the action~(\ref{3finf}),
one arrives at $\alpha_{ij}=0$ and
\begin{eqnarray}
\phi''+2\cH \phi'+12a^2 V_I\phi =0,\label{bg-ph}
\end{eqnarray}
where $I$ for the background is given by $I=6\phi^2$.
Thus, the field $\phi$ evolves according to the same equation of motion
as in conventional scalar field inflation.
The energy-momentum tensor is found to be
$T_0^{\;0}=-(\phi')^2/2a^2-V$ and $T_i^{~j}=[(\phi')^2/2a^2-V]\delta_i^{~j}$,
implying that the 3-form field is indeed compatible with isotropy.
We would like to stress that the non-minimal coupling of the 3-form to gravity
is essential to obtain the desired inflationary background similar to the standard one.

As is almost clear from the above derivation,
2- and 3-form inflation models have their dual description in terms of
1- and 0-forms.
However, they are not equivalent to vector and standard scalar field inflation models.
In the dual description, they correspond to vector and scalar field theories
with non-minimal kinetic terms~\cite{p-form}. In this sense, $p$-forms provide novel inflationary models.

\section{Action for gravitational waves}

Let us consider tensor perturbations
$h_{ij}=h_{ij}(\eta, \mathbf{x})$ on a $p$-form inflationary background:
\begin{eqnarray}
\D s^2=a^2(\eta)\left[-\D \eta^2+(\delta_{ij}+h_{ij})\D x^i\D x^j\right],
\end{eqnarray}
where $h_{ij}$ is transverse and traceless, $h_i^{~i}=\partial_jh_i^{~j}=0$.
In the standard inflationary scenarios driven by one or more scalar fields,
the behavior of gravitational waves is completely determined by
the background geometry, and hence identical expansion histories
give the identical evolution of gravitational waves.
In the case of $p$-form inflation, however, the situation is more involved,
and $p$-form inflationary models predict
different evolution of gravitational waves than that in
the corresponding scalar field inflation.
The 1- and 2-form cases are particularly difficult to analyze in general, because
in contrast to the standard linear perturbation theory,
scalar, vector, and tensor modes are coupled due to the presence of the background form fields.
This point is however circumvented by considering a large number of randomly oriented fields
that suppress the couplings statistically~\cite{V2}.
For example, terms like $B_i\delta B_j$ do not contribute to the equation of motion at leading order.
Therefore, we may separate the evolution of gravitational waves from
the contributions of vector and scalar perturbations.
The behavior of gravitational waves from vector inflation
was studied in~\cite{V2}.
In the present paper, we generalize the analysis of~\cite{V2} to 2- and 3-form inflation.

In order to obtain the action for gravitational waves,
we must expand Eqs.~(\ref{2fac}) and~(\ref{3finf}) to second order in $h_{ij}$.
Since lengthy calculations need to be done for this, the detailed derivation
is presented in Appendix~\ref{app:detail}.
Here we only provide the final result.

\subsection{2-form inflation}

The action for the gravitational waves from 2-form inflation is given by
\begin{eqnarray}
S_2 \approx\int  \frac{a^2}{8\kappa^2} 
\left[
\left(h_{ij}'\right)^2-c_s^2\left(\partial_k h_{ij}\right)^2-m_g^2h_{ij}^2
\right]\D\eta\,\D^3 x,\label{2ndaction-2f}
\end{eqnarray}
where
\begin{eqnarray}
c_s^2:=1- \frac{2}{3}\kappa^2 NB^2
\end{eqnarray}
is the propagation speed of the gravitational waves and
\begin{eqnarray}
m_g^2:= \frac{4\kappa^2N}{3} \left[
 4V_Ia^2B^2+\frac{8}{5}V_{II}a^2B^4-\frac{a''}{a}B^2-
\left( B'+ \cH B\right)^2\right]\label{mass-2}
\end{eqnarray}
is the graviton mass.
The equation of motion can be derived from the action~(\ref{2ndaction-2f}),
or, directly from the linearized Einstein equations:
\begin{eqnarray}
h_{ij}''+2\cH h_{ij}'-c_s^2\nabla^2h_{ij}=-m_g^2h_{ij}.
\end{eqnarray}

Let us, for example, consider a simple chaotic potential $V=m^2I/4$. In this case,
we may use the approximation
$3\cH^2\simeq \kappa^2Na^2V$,
$a''/a^3=\cH'+\cH^2\simeq 2\cH^2-\kappa^2N(B')^2/2$, and
$(B'+\cH B)^2\simeq a^2H^2B^2(1-2m^2/3H^2)$. Then, Eq.~(\ref{mass-2})
is simplified to
\begin{eqnarray}
m_g^2\simeq 4\cH^2\left(4-\kappa^2NB^2\right).\label{mass-2-c}
\end{eqnarray}
The similar graviton mass term also arises in vector inflation~\cite{V2}.
If $\kappa^2NB^2>4$, we have the large tachyonic mass,
$m_g^2 < 0$, with $m_g^2\sim{\cal O}( \cH^2)$,
implying the unstable evolution of tensor perturbations.
In vector inflation, $\kappa^2NB^2\gtrsim 1$ is required in order for chaotic inflation to take place~\cite{V1}.
This is also the case for 2-form inflation.
Therefore, it is difficult to realize the large field inflationary models
within the context of 2-form inflation.

The most striking nature of 2-form inflation appears in the
propagation speed of the gravitational waves.
In vector inflation, one finds the usual propagation speed, i.e., $c_s^2=1$~\cite{V2}.
In 2-form inflation, however, $c_s^2$ depends on the background field value
and hence differs from unity in general. In particular,
if $\kappa^2NB^2>3/2$, the propagation speed squared becomes negative, which is pathological.
This shows that, in addition to the above mentioned tachyonic mass,
2-form inflation suffers from the negative sound speed squared
in the case of the large field models.
The varying speed of gravitational wave propagation also arises in~\cite{Gasperini:1997up}.

\subsection{3-form inflation}

The action for the gravitational waves from 3-form inflation is given by
\begin{eqnarray}
S_3 =\int \frac{a^2}{8\kappa^2}\Omega^2
\left[
\left(h_{ij}'\right)^2-c_s^2\left(\partial_k h_{ij}\right)^2\right]\D\eta\,\D^3 x,
\end{eqnarray}
where
\begin{eqnarray}
\Omega^2:=1+\frac{3}{2}\kappa^2\phi^2
\end{eqnarray}
and
\begin{eqnarray}
c_s^2:=\frac{2-\kappa^2\phi^2}{2+3\kappa^2\phi^2}.\label{cs2_3f}
\end{eqnarray}
The equation of motion is
\begin{eqnarray}
h_{ij}''+ 2\left(\cH +\frac{\Omega'}{\Omega}\right)h_{ij}'-c_s^2\nabla^2h_{ij}=0.
\end{eqnarray}

Contrary to 1- and 2-form inflation, 3-form inflation does not give rise to
the graviton mass term, and so in this case one does not need to worry about the
tachyonic mass.
However, 3-form inflation has the varying speed of propagation
of gravitational waves, as in 2-form inflation.
As is clear from Eq.~(\ref{cs2_3f}), one obtains $c_s^2<0$ for $\kappa \phi >\sqrt{2}$,
which basically rules out the large field models.

\section{Generation of gravitational waves from p-form inflation}

\begin{table}[t]
 \caption{The graviton mass $m_g^2$, the propagation speed of gravitational waves
 $c_s^2$, and the prefactor $\Omega^2$ in $p$-form inflation for different $p$.
 For completeness we include the cases with $p=0$ (standard scalar field inflation)
 and $p=1$ (vector inflation).}%
 \begin{center}
  \begin{tabular}{|c|c|c|c|}
    \hline
       & $m_g^2$   & $c_s^2$   & $\Omega^2$   \\
           \hline
$p=0$       & 0 &   1 &  1  \\
    \hline
$p=1$       &  Eq.~(6) of Ref.~\cite{V2} &   1 &  $\displaystyle{1+\frac{1}{6}\kappa^2NB^2}$  \\
    \hline
$p=2$       &  Eq.~(\ref{mass-2})  & $\displaystyle{1-\frac{2}{3}\kappa^2NB^2}$   & 1   \\
    \hline
$p=3$       & 0   &  $\displaystyle{\frac{2-\kappa^2\phi^2}{2+3\kappa^2\phi^2}}$  & $\displaystyle{1+\frac{3}{2}\kappa^2\phi^2}$   \\
    \hline
  \end{tabular}
 \end{center}
\end{table}

In this section, we quantize the actions for the $p$-form inflationary gravitational waves to
discuss the power spectrum,
following Refs.~\cite{V2} and \cite{SL}. 

The tensor perturbation is expanded into Fourier modes as
\begin{eqnarray}
h_{ij} (\eta, \mathbf{x}) = \int
 {\D^3 k \over (2\pi)^{3/2}}h_{\mathbf{k}}(\eta) e_{ij}(\mathbf{k}) \rm{e}^{i\mathbf{k}\mathbf{x}}~,
\end{eqnarray}
where $e_{ij}(\mathbf{k})$ is the polarization tensor.
The spectrum of gravitational waves is commonly defined by
\begin{eqnarray}
\langle h_{\mathbf{k}}h_{\mathbf{k}'}\rangle \equiv {2\pi^2 \over k^3}{\cal P}_T(k)\delta^{(3)}
\left(\mathbf{k} + \mathbf{k}' \right)~.
\end{eqnarray}
In terms of a new variable and a new time coordinate defined by
\begin{eqnarray}
&& v_{\mathbf{k}} := z h_{\mathbf{k}} ~,~ z := {a\sqrt{c_s} \Omega \over 2}~, \\
&& \D y := c_s \D \eta~, 
\end{eqnarray}
the action can be rewritten as
\begin{eqnarray}
S_{{\rm GW}} = {1 \over 2\kappa^2}\int \D^3k\,\D y  
\left\{ \left|v_{\mathbf{k},y}\right|^2 - 
\left[k^2 -\left({z_{,yy} \over z} - {m_g^2 \over c_s^2} \right)\right]
\left|v_{\mathbf{k}}\right|^2 \right\},
\end{eqnarray}
where $m_g^2$, $c_s^2$, and $\Omega^2$ can be found in Table~I.
Then, the equation of motion for $v_\mathbf{k}$ reduces to
\begin{eqnarray}
v_{,yy} +\left[k^2 - \left(1+\alpha\right)\frac{a_{,yy}}{a} \right] v = 0,
\end{eqnarray}
where we have abbreviated a suffix $\mathbf{k}$ and introduced a parameter $\alpha$
to describe the deviation from the well-known formula for
the gravitational waves from the standard scalar field inflation model.\footnote{In
the limiting case with $H,\,c_s,\,\Omega,\,\alpha =$ const.,
we have an exact solution
\begin{eqnarray*}
v = \frac{\sqrt{\pi}}{2}{\rm e}^{i(2\nu+1)\pi/4}aHc_s^{1/2}(-\eta)^{3/2}H^{(1)}_\nu(-c_s k\eta),
\quad \nu:=\frac{3}{2}\left(1+\frac{8}{9}\alpha\right)^{1/2},
\end{eqnarray*}
where $H_\nu^{(1)}$ is the Hankel function of the first kind of order $\nu$.
}
The appropriate initial condition is given by
\begin{eqnarray}
v \to {1 \over \sqrt{2k}}e^{-iky}\quad\text{for}\quad
ky \to -\infty \quad\text{(the Bunch-Davies vacuum)}.
\end{eqnarray}

It is useful to express $\alpha$ in terms of slow-roll parameters.
We define the slow-roll parameters as
\begin{eqnarray}
\epsilon:= 1-\frac{\cH'}{\cH^2},
\quad
s:= \frac{c_s'}{\cH c_s},
\quad
\omega:=\frac{\Omega'}{\cH\Omega},
\end{eqnarray}
with the slow-roll condition $\epsilon\ll 1$.
Since $c_s$ and $\Omega$ are functions of the field,
they may be thought of as slowly varying functions of time.
We assume that $s,\,\omega\lesssim{\cal O}(\epsilon^{1/2})$,
and the estimate will be verified later.
Using these slow-roll parameters, the deviation parameter $\alpha$ can be written as
\begin{eqnarray}
\alpha = \alpha_1 + \alpha_2
\end{eqnarray}
with
\begin{eqnarray}
\alpha_1 &=& {3 \over 4}s+{3 \over 2}\omega+{1 \over 4}{s' \over \cH}
+{1 \over 2}{\omega' \over \cH} + {1 \over 4}s^2 + {1 \over 2}\omega^2
+{3 \over 4}s \omega+{\cal  O}(\epsilon^{3/2}),
\\
\alpha_2 &=& 
-{m_g^2 \over 2\cH^2}\left(1 - {1 \over 2}\epsilon - {1 \over 2}s \right)^{-1}~,
\end{eqnarray}
where $\alpha_1$ represents the correction from the time variation of $c_s$ and $\Omega$,
and $\alpha_2$ the correction from the graviton mass.

In the limit where the slow-roll conditions for $\epsilon$, $s$, and $\omega$
are satisfied and the graviton mass is small,
we have $\alpha \approx 0$, leading approximately to the scale invariant spectrum of gravitational waves:
\begin{eqnarray}
{\cal P}_T^{1/ 2} \sim {\kappa H \over 2 \pi \Omega c_s^{3/2}}\biggr|_{aH=c_s k}.
\end{eqnarray}
If the graviton mass squared is large and positive, i.e, $m_g^2\sim{\cal O}(\cH^2 )>0$,
the growth of the tensor perturbations is suppressed.
If, on the other hand, the mass squared is large and negative, then we have
the unstable evolution of tensor perturbations. 
As mentioned in the previous section, 
the mass squared clearly becomes large and negative in chaotic inflation.
In addition to this problem,
we have to be careful about the negative sound speed squared in the large field models
of 2- and 3-form inflation,
which is pathological.
For this reason, in what follows
we focus on the cases with $c_s^2>0$
and evaluate in more detail the correction term $\alpha$
which gives the scale dependence of the spectrum.


\subsection{2-form inflation}

In the case of 2-form inflation, we explicitly have
\begin{eqnarray}
\epsilon=\frac{1}{2\kappa^2N}\left(\frac{4V_IB}{V}\right)^2,
\quad
s = \sigma (2\epsilon)^{1/2}\frac{2\beta}{3-2\beta^2},
\quad\omega=0,
\end{eqnarray}
where $\beta:=\kappa\sqrt{N}B$ and
$\sigma=1$ (respectively $\sigma=-1$) for $V_IB/V>0$ (respectively $V_IB/V<0$).
Now it is easy to see $s\lesssim{\cal O}(\epsilon^{1/2})$.
The graviton mass reduces to
\begin{eqnarray}
\frac{m_g^2}{4\cH^2}=
-\beta^2\left(1-\frac{\epsilon}{3}\right)-\frac{2\epsilon}{3}+\frac{5}{3}\sigma(2\epsilon)^{1/2}\beta
+\frac{8}{5}\frac{V_{II}}{V}B^4.\label{mass_explicit}
\end{eqnarray}
In deriving the above equations we have used the approximation
$3 \cH^2 \simeq \kappa^2 N a^2 V$ and $3\cH B' \simeq - 4a^2 V_I B$.
It is straightforward to check that
for the chaotic potential $V=m^2I/4$ Eq.~(\ref{mass_explicit}) reproduces Eq.~(\ref{mass-2-c}). 
One sees that the leading correction is given by
\begin{eqnarray}
\alpha\sim\text{max}\left\{
\epsilon,\, \beta^2,\, \epsilon^{1/2}\beta,\,\frac{V_{II}}{V}B^4
\right\}.
\end{eqnarray}
In principle we can take $\beta\sim{\cal O}(1)$ while keeping $c_s^2>0$, so that
the correction is large.
However, this case seems irrelevant because we have $m_g^2\approx-4\cH^2<0$
unless the last term in Eq.~(\ref{mass_explicit}) is fine-tuned to cancel this negative contribution.

\subsection{3-form inflation}

Let us next consider 3-form inflation.
Using $3 \cH^2 \simeq \kappa^2 a^2 V$ and $3\cH \phi' \simeq -12a^2V_I\phi$,
one finds
\begin{eqnarray}
\epsilon=\frac{1}{2\kappa^2}\left(\frac{12V_I\phi}{V}\right)^2,
\quad
s=\sigma(2\epsilon)^{1/2}\frac{8\beta}{(2-\beta^2)(2+3\beta^2)},
\quad
\omega=-\sigma(2\epsilon)^{1/2}\frac{3\beta}{2+3\beta^2},
\end{eqnarray}
where $\beta:=\kappa\phi$ and
$\sigma=1$ (respectively $\sigma=-1$) for $V_I\phi/V>0$ (respectively $V_I\phi/V<0$).
One can verify $s,\,\omega \lesssim{\cal O}(\epsilon^{1/2})$.
In the small field models ($\kappa\phi\ll 1$), the leading correction is given by
\begin{eqnarray}
\alpha\simeq-\frac{3}{4}\sigma(2\epsilon)^{1/2}\kappa\phi.
\end{eqnarray}
With a relatively large field value $\kappa\phi\sim 1$ and positive $c_s^2$,
the correction could be as large as ${\cal O}(\epsilon^{1/2})$.

\section{Conclusions}

We have studied gravitational waves generated during $p$-form inflation.
In the case of 2-form inflation, we considered a large number of 2-form fields
so that the model is compatible with the background isotropy~\cite{V1, p-form}.
The main feature of gravitational waves from 2-form inflation can be found
in the mass term, $m_g^2$, and the propagation speed, $c_s^2$.
We obtained the mass term very similar to that in vector inflation.
The mass squared becomes negative in large field models,
as in vector inflation, implying the unstable evolution of gravitational waves.
In addition to this,
2-form inflation predicts the varying speed of gravitational wave propagation, and
the large field models are found to give $c_s^2<0$. Thus, the large field models
of 2-form inflation are unlikely to be viable.
In contrast to 1- and 2-forms, a 3-form is compatible with isotropy.
With some particular coupling to gravity, the background evolution of 3-form inflation is 
very similar to that driven by a scalar field.
However, the behavior of gravitational waves is different again.
Although 3-form inflation does not give rise to the mass term of gravitons,
we showed that the propagation speed of gravitational waves differs from unity
also in 3-form inflation.
As the squared speed becomes negative when the field value is large,
it is difficult to construct working large field models in the context of 3-form inflation.
We also showed that the correction to the spectrum of gravitational waves
is very small in the small field models of 2- and 3-form inflation.

Finally, we would like to remark that in a spatially curved universe
the background evolution of $p$-form inflation will be different from
the corresponding scalar field case.
The non-zero spatial curvature indeed affects the onset of 1-form inflation~\cite{Chiba}.
It would be interesting to study the dynamics of general $p$-form inflation
in the case of a spatially curved universe.

\acknowledgments
T.K. is supported by the JSPS under Contact No.~19-4199.
S.Y. is supported in part by Grant-in-Aid for Scientific Research
on Priority Areas No. 467 ``Probing the Dark Energy through an
Extremely Wide and Deep Survey with Subaru Telescope''.
He also acknowledges the support from the Grand-in-Aid for the Global COE Program
``Quest for Fundamental Principles in the Universe: from Particles to the Solar
 System and the Cosmos '' from the
Ministry of Education, Culture, Sports, Science and Technology (MEXT) of
Japan. 

\subsection*{Note added}
A day after this paper appeared on arXiv, a related paper~\cite{V3} has also appeared,
in which the authors study cosmological perturbations from vector inflation.

\appendix

\section{Calculation details}\label{app:detail}

\subsection{The metric and Ricci tensor}

The perturbed metric we consider is
\begin{eqnarray}
\D s^2=a^2(\eta)\left[-\D \eta^2+(\delta_{ij}+h_{ij})\D x^i\D x^j\right],
\end{eqnarray}
where $\delta^{ij}h_{ij}=\partial^jh_{ij}=0$. The inverse metric is then given by
\begin{eqnarray}
g^{ij}=a^{-2}\left(\delta^{ij}-h^{ij}+h^{ik}h_{k}^{\;j}\right).
\end{eqnarray}
For this metric we have
\begin{eqnarray}
\sqrt{-g}&=&a^4\left(1-\frac{1}{4}h_{ij}^2\right),
\\
R_{00}&=&-3\cH'+\frac{1}{2}h_{ij}h_{ij}''+\frac{1}{4}h_{ij}'h_{ij}'+\frac{1}{2}\cH h_{ij}h_{ij}',
\\
R_{ij}&=&\left(\cH'+2\cH^2\right)\left(\delta_{ij}+h_{ij}\right)
+\frac{1}{2}\left(h_{ij}''+2\cH h_{ij}'-\nabla^2h_{ij}\right)
-\frac{1}{2}h_{ik}'h_{jk}'-\frac{1}{2}\cH\delta_{ij}h_{kl}h_{kl}'
\nonumber\\&&\quad
+\frac{1}{4}\partial_i h_{kl}\partial_jh_{kl} 
+\frac{1}{2} h_{kl}\partial_i\partial_jh_{kl}
-\frac{1}{2}h_{kl}\partial_k\left(
\partial_ih_{jl}+\partial_jh_{il}-\partial_lh_{ij}
\right)
-\frac{1}{2}\partial_kh_{il}\partial_lh_{jk}+\frac{1}{2}\partial_kh_{il}\partial_kh_{jl}
,
\\
R&=&\frac{1}{a^2}\left[
6\frac{a''}{a}-h_{ij}h_{ij}''-\frac{3}{4}h_{ij}'h_{ij}'-3\cH h_{ij}h_{ij}'+h_{ij}\nabla^2h_{ij}
+\frac{3}{4}\left(\partial_kh_{ij}\right)^2-\frac{1}{2}\partial_k\left(h_{ij}\partial_ih_{jk}\right)
\right],
\end{eqnarray}
leading to
\begin{eqnarray}
\sqrt{-g}R= \frac{a^2}{4}\left[
(h_{ij}')^2-\left(\partial_kh_{ij}\right)^2
\right]-\frac{a^2}{2}\left(\cH^2+2\cH' \right)h_{ij}^2 ,
\label{rgr}
\end{eqnarray}
where summation over repeated indices is understood.
Note that a total derivative term is omitted in Eq.~(\ref{rgr}).

\subsection{The 2-form}

We start with computing $I^{(a)}:=A_{\mu\nu}^{(a)2} $.
Each $I$ is explicitly given by
\begin{eqnarray}
I&=&b_{ij}^2-2h_{ij}b_{ik}b_{jk}+h_{ij}h_{kl}b_{ik}b_{jl}+2h_{ik}h_{jk}b_{il}b_{jl}
\nonumber\\
&=&2B_i^2+2h_{ij}B_iB_j+h_{ij}h_{kl}\varepsilon_{ikm}\varepsilon_{jln}B_mB_n
+2h_{ij}^2B_k^2-2h_{ik}h_{jk}B_iB_j,
\end{eqnarray}
so that
\begin{eqnarray}
\sum_{a=1}^{N}I^{(a)} \approx2NB^2+NB^2h_{ij}^2,
\end{eqnarray}
where we used Eq.~(\ref{form_stat}).
The field strength is $F_{0ij}=a^2\varepsilon_{ijk}\left(B_k'+2\cH B_k\right)$,
and the kinetic term of the 2-form field is simply given by
\begin{eqnarray}
\sqrt{-g}\sum_{a=1}^{N}\frac{1}{12}F_{\mu\nu\rho}^{(a)2}
\approx-\frac{Na^2}{2}\left(B'+2\cH B\right)^2\left(1+\frac{1}{4}h_{ij}^2\right),
\end{eqnarray}
while the potential term reduces to
\begin{eqnarray}
\sqrt{-g}\sum_{a=1}^{N}V(I^{(a)}) \approx 
a^4 \left[
NV+ \left( - {N \over 4}V + NV_IB^2 +\frac{4N}{15}V_{II}B^4
\right)
h_{ij}^2\right],
\end{eqnarray}
where we used the formula
\begin{eqnarray}
\sum_{a=1}^{N}B_i^{(a)}B_j^{(a)}B_k^{(a)}B_l^{(a)}h_{ij}h_{kl}
\approx \frac{2}{15}NB^4h_{ij}^2.
\end{eqnarray}
Finally, the coupling terms are
\begin{eqnarray}
\sum_{a=1}^{N}\sqrt{-g}\frac{1}{6}RI^{(a)}&\approx&
2Naa''B^2
+
\frac{Na^2B^2}{12} \left[\left(h_{ij}'\right)^2-\left(\partial_kh_{ij}\right)^2\right]
\nonumber\\&&\qquad
+\frac{Na^2}{3}\left[
2\frac{a''}{a}B^2-(B')^2+\frac{1}{2}\cH^2B^2+\cH BB'+4a^2V_IB^2
\right]h_{ij}^2,
\end{eqnarray}
and
\begin{eqnarray}
\sum_{a=1}^N\frac{1}{2}\sqrt{-g}A_{\mu\nu}^{(a)}R^{\nu}_{~\rho }A^{\rho\mu}_{(a)}
&\approx& -Na^2B^2\left(\cH'+2\cH^2\right)
-\frac{3}{4}Na^2B^2\left(\cH'+2\cH^2\right)h_{ij}^2
\nonumber\\&&\qquad
+\frac{1}{2}Na^2B^2h_{ij}\delta R_{ij}^{(1)}
-\frac{1}{3}Na^2B^2\delta_{ij}\delta R_{ij}^{(2)},
\end{eqnarray}
with
\begin{eqnarray}
Na^2B^2 h_{ij}\delta R_{ij}^{(1)}=
-\frac{1}{2}Na^2B^2\left[\left(h_{ij}'\right)^2-\left(\partial_kh_{ij}\right)^2\right]
+Na^2\left[
\frac{1}{2}(B')^2-2a^2B^2V_I+\left(\frac{a''}{a}+\cH^2\right)B^2
\right]h_{ij}^2,
\end{eqnarray}
and
\begin{eqnarray}
Na^2B^2 \delta_{ij}\delta R_{ij}^{(2)}=
Na^2B^2\left[
-\frac{1}{2}\left(h_{ij}'\right)^2+\frac{1}{4}\left(\partial_kh_{ij}\right)^2
\right]
+\frac{3}{4}Na^2\left[
2\cH BB'+\left(\frac{a''}{a}+\cH^2\right)B^2
\right]h_{ij}^2,
\end{eqnarray}
where we used the background equation~(\ref{bg:bi}) and
integration by parts, and removed total derivative terms.

\subsection{The 3-form}

$I:=A_{\mu\nu\rho}^2$ is given by
\begin{eqnarray}
I = 6\phi^2\left(1+\frac{1}{2}h_{ij}^2\right).
\end{eqnarray}
The field strength is $F_{0ijk}=a^3\left(\phi'+3\cH \phi\right)\varepsilon_{ijk}$, and
the kinetic term of the 3-form field is
\begin{eqnarray}
\sqrt{-g}\frac{1}{48}F^2_{\mu\nu\rho\sigma}
=-\frac{1}{2}a^2\left(\phi'+3\cH \phi\right)^2\left(1+\frac{1}{4} h_{ij}^2\right).
\end{eqnarray}
The potential term is
\begin{eqnarray}
\sqrt{-g}V(I) = a^4\left[V +\left(-\frac{1}{4}V+V_I \cdot3\phi^2\right) h_{ij} ^2\right].
\end{eqnarray}
The first one of the coupling terms is
\begin{eqnarray}
\frac{1}{8}\sqrt{-g}RI=\frac{9}{2}aa''\phi^2+
\frac{3a^2\phi^2}{16} \left[(h_{ij}')^2-\partial_k h_{ij}^2\right]+
\frac{3a^2}{4}\left\{
2\frac{a''}{a}\phi^2-(\phi')^2+\frac{1}{2}\cH^2\phi^2+\cH \phi\phi'+12a^2V_I\phi^2
\right\}h_{ij}^2.
\end{eqnarray}
Noting that
\begin{eqnarray}
A_{\mu\nu}^{~~~ i }A^{j\mu\nu}=2 \phi^2g^{ij}+
a^{-2}\phi^2\left[
2\left(\delta_{ij}h_{kl}^2-h_{ik}h_{jk}\right)+\varepsilon_{ikm}\varepsilon_{jln}h_{kl}h_{mn}
\right],
\end{eqnarray}
we get
\begin{eqnarray}
\sqrt{-g}A_{\mu\nu\rho}R^{\rho\sigma}A_{\sigma}^{~\mu\nu}&=&
6a^2\phi^2\left(\frac{a''}{a}+\cH^2\right)
-\frac{1}{2}a^2\phi^2(\partial_k h_{ij})^2
\nonumber\\&&\qquad
+a^2\left\{
3\phi^2\left(\frac{a''}{a}+\cH^2\right)+3\phi\phi'\cH-(\phi')^2
+12a^2\phi^2V_I
\right\}h_{ij}^2.
\end{eqnarray}
In computing the coupling terms we used the background equation~(\ref{bg-ph})
and integration by parts, and removed total derivative terms.


\end{document}